\documentclass{PoS}

\newcommand{\be}{\begin{equation}}
\newcommand{\ee}{\end{equation}}

\usepackage{multirow}
\usepackage{amsmath}
\usepackage{tabularx}
\usepackage{booktabs} 
\usepackage{caption} 

\title{Non-ordinary nature of the $f_0(500)$ resonance from its Regge trajectory}

\ShortTitle{Non-ordinary nature of the $f_0(500)$ resonance from its Regge trajectory}

\author{\speaker{J.~Nebreda}\\
        Yukawa Institute for Theoretical Physics, Kyoto University, Kyoto, 606-8502, Japan\\
        Center for Exploration of Energy and Matter, Indiana University, Bloomington, IN 47403, USA\\
        Physics Department  Indiana University, Bloomington, IN 47405, USA\\
        Dpto. de F\'isica Te\'orica II, Universidad Complutense de Madrid, 28040, Spain
        E-mail: \email{jnebreda@yukawa.kyoto-u.ac.jp}}
        
\author{J.~T.~Londergan\\        
Center for Exploration of Energy and Matter, Indiana University, Bloomington, IN 47403, USA\\
 Physics Department  Indiana University, Bloomington, IN 47405, USA}

\author{J.~R.~Pelaez\\        
 Dpto. de F\'isica Te\'orica II, Universidad Complutense de Madrid, Madrid, 28040, Spain
}

\author{A.~P.~Szczepaniak\\        
 Center for Exploration of Energy and Matter, Indiana University, Bloomington, IN 47403, USA\\
 Physics Department  Indiana University, Bloomington, IN 47405, USA\\
 Jefferson Laboratory, 12000 Jefferson Avenue, Newport News, VA 23606, USA}

\abstract{
We report here our results~\cite{ours} on how to obtain the Regge trajectory of a resonance from its pole in a
scattering process by imposing analytic constraints in the complex angular momentum plane.
 The method, suited for resonances that dominate an elastic
scattering amplitude, has been applied to the $\rho(770)$ and the $f_0(500)$ resonances.
Whereas for the former we obtain a linear Regge trajectory, characteristic of ordinary
quark-antiquark states, for the latter we find a non-linear trajectory with a much
smaller slope at the resonance mass. Moreover, we show that if a linear trajectory with a
slope of typical size is imposed for the $f_0(500)$, the corresponding amplitude is at
odds with the data. This provides a strong indication of the non-ordinary nature of the
sigma meson.}

\FullConference{XV International Conference on Hadron Spectroscopy-Hadron 2013\\
		4-8 November 2013\\
		Nara, Japan }

\begin{document}


\section{Introduction}\label{aba:sec1}

 The analytical properties of amplitudes in the complex angular momentum plane allow us to investigate 
  the dynamical linkage of resonances of different spins. The function connecting such 
 resonances is known as the Regge trajectory and its form can be used to discriminate between the underlying (QCD) mechanisms responsible for generating the resonances. In particular,  linear $(J,M^2)$
trajectories relating the angular momentum $J$ and the mass squared are naively and intuitively interpreted in terms of the rotation 
of the flux tube connecting a quark and an antiquark. Strong deviations from this linear behavior would suggest a rather different nature.

We have applied this method to the lightest resonances in elastic $\pi\pi$ scattering: the $\rho(770)$, which suits well the 
ordinary meson picture, and the $f_0(500)$ or $\sigma$ meson, whose nature is still the subject of a longstanding debate and which does not seem to fit well in the $(J,M^2)$ trajectories \cite{Anisovich:2000kxa}. 

\section{Regge trajectory from a resonance pole}

An elastic $\pi\pi$ partial wave near a Regge pole reads
\be
t_l(s)  = \beta(s)/(l-\alpha(s)) + f(l,s),
\label{Reggeliket}
\ee
where $f(l,s)$ is a regular function of $l$, and the Regge trajectory $\alpha(s)$ and 
residue $\beta(s)$ are analytic functions, the former having a cut along the real axis for $s > 4m_\pi^2$. Now, if the pole dominates in Eq.\eqref{Reggeliket}, the unitarity condition implies that, for real $l$, 
\be
\mbox{Im}\,\alpha(s)   = \rho(s) \beta(s).   \label{unit} 
\ee
 
 On the other hand, taking into account the threshold behavior and making explicit the cancellation of the poles of the Legendre function appearing in the full amplitude, we can write the $\beta(s)$ function as~\cite{Chu:1969ga}
\be
\beta(s) =  \gamma(s) \hat s^{\alpha(s)} /\Gamma(\alpha(s) + 3/2) , \label{reduced} 
\ee
where $\hat s =( s-4m_\pi^2)/s_0$. The dimensional scale $s_0=1\,$ GeV$^2$ is introduced for 
convenience and the reduced residue $\gamma(s)$ is an analytic function, whose phase is known because $\beta(s)$ is real in the real axis.

We can thus write down dispersion relations for $\alpha(s)$ and $\beta(s)$, and connect them by using the unitarity condition. This way we obtain the following system of integral equations~\cite{Chu:1969ga}:
\begin{align}
\mbox{Re}\, \alpha(s) & =   \alpha_0 + \alpha' s +  \frac{s}{\pi} PV \int_{4m_\pi^2}^\infty ds' \frac{ \mbox{Im}\,\alpha(s')}{s' (s' -s)}, \label{iteration1}\\
\mbox{Im}\,\alpha(s)&=  \frac{ \rho(s)  b_0 \hat s^{\alpha_0 + \alpha' s} }{|\Gamma(\alpha(s) + \frac{3}{2})|}
 \exp\Bigg( - \alpha' s[1-\log(\alpha' s_0)]\\
&+ \frac{s}{\pi} PV\!\int_{4m_\pi^2}^\infty\!\!ds' \frac{ \mbox{Im}\,\alpha(s') \log\frac{\hat s}{\hat s'} + \mbox{arg }\Gamma\left(\alpha(s')+\frac{3}{2}\right)}{s' (s' - s)} \Bigg), 
\label{iteration2}\\
 \label{betafromalpha}
 \end{align}
where $PV$ denotes ``principal value'' and $\alpha_0, \alpha'$ and $b_0$ are free parameters that need to be determined phenomenologically.  

For the $\sigma$-meson, $\beta(s)$ at low energies should also
include the Adler-zero required by chiral symmetry. 
In practice it is enough to multiply
the right hand side of Eq.\eqref{iteration2} by $2s-m_\pi^2$ (Adler zero at leading order in Chiral Perturbation Theory~\cite{chpt}) and replace the $3/2$ by $5/2$ inside the gamma functions in order not to spoil the large $s$-behavior. Note that $b_0$ now is not dimensionless.

\section{$\rho(770)$ and $f_0(500)$ Regge trajectories}
  
\begin{figure}
\centering
\includegraphics[scale=0.60,angle=-90]{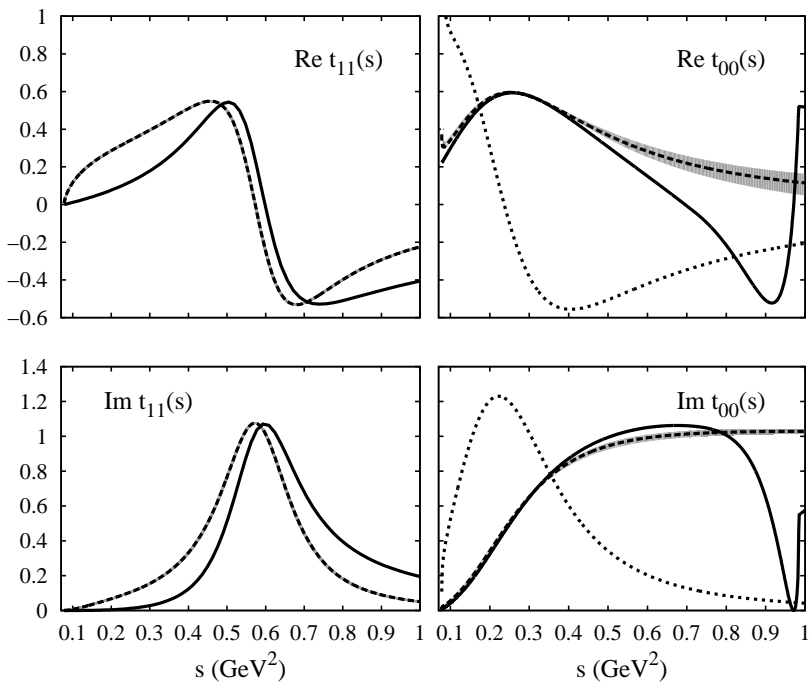}
 \caption{\rm \label{fig:ampl} 
  Partial waves $t_{lI}$ with $l=1$ (left panels) and $l=0$ (right panels). 
  Solid lines represent the amplitudes from \protect{\cite{GarciaMartin:2011jx}}. Their corresponding resonance poles 
  are the input for the constrained Regge-pole amplitudes shown with dashed curves. The gray bands cover the uncertainties due to the errors in the determination of the pole positions and residues from 
 the dispersive analysis of data in \cite{GarciaMartin:2011jx}. In the right panels, the dotted lines represent the constrained Regge-pole amplitude for the $S$-wave 
if the $\sigma$-pole is fitted by  imposing a linear trajectory with $\alpha'\simeq 1\,$GeV$^{-2}$.}
\end{figure}

For a given set of $\alpha_0, \alpha'$ and $b_0$ parameters we solve the system of Eqs.~\eqref{iteration1} and \eqref{iteration2} iteratively. From only three inputs, namely, the real and imaginary parts of the resonance pole position $s_M$ and the absolute value of the residue $|g_M|$, we can determine the $\alpha_0, \alpha',b_0$ parameters of the corresponding Regge trajectories,  by requiring that at the pole, on the second Riemann sheet,
$\beta_M(s)/(l  - \alpha_M(s))\rightarrow |g^2_M|/(s-s_M)$, 
with $l=0,1$ for $M=\sigma,\rho$. The pole parameters are taken from a precise dispersive representation of $\pi\pi$ scattering data \cite{GarciaMartin:2011jx}. They are shown in Table \ref{aba:tbl1} together with the corresponding output values. In Fig.~\ref{fig:ampl}  we compare the obtained Regge amplitude on the real axis with the partial waves of \cite{GarciaMartin:2011jx}. Let us point out that they do not need to overlap since they are only constrained to agree at the resonance pole. However, we find a fair agreement in the resonant region, which, as expected,  deteriorates as we approach threshold or the inelastic region, specially in the case of the $S$-wave due to the interference with the $f_0(980)$.

\vspace{6mm}
\begin{minipage}{\linewidth}
\centering
\captionof{table}{Input and output resonance pole positions and residues} \label{aba:tbl1} 
\begin{tabular}{@{}ccccc@{}}\toprule
& \multicolumn{2}{c}{Input} & \multicolumn{2}{c}{Output} \\
& $\sqrt{s_M}$ (MeV) & $|g_M|$ & $\sqrt{s_M}$ (MeV) & $|g_M|$ \\\midrule
$\rho(770)$ & $763.7^{+1.7}_{-1.5}-i73.2^{+1.0}_{-1.1}$ & $6.01^{+0.04}_{-0.07}$ & $762.7-i73.5$ & 5.99 \\
$f_0(500)$ & $457^{+14}_{-13}-i279^{+11}_{-7}$ & $3.59^{+0.11}_{-0.13}$ GeV & $461-i281$ & $3.51$ GeV \\
\bottomrule
\end{tabular}\par
\bigskip
\end{minipage}
\vspace{4mm}

\begin{figure}
\centering
\hspace{-5mm}\includegraphics[scale=0.55,angle=-90]{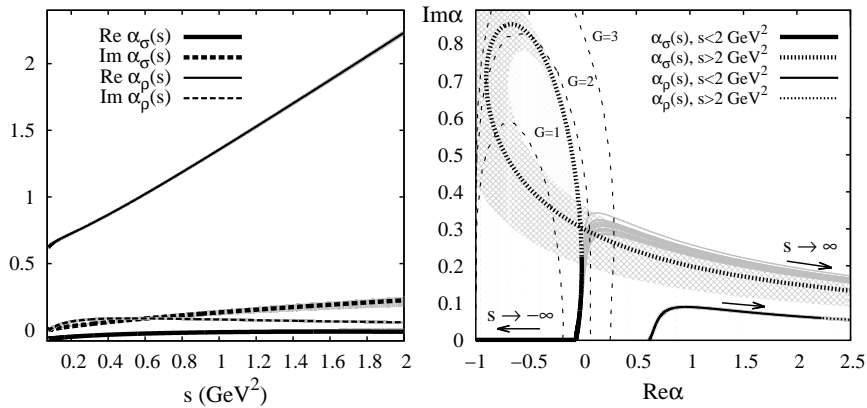}
 \caption{\rm \label{fig:trajectories} 
  (Left) $\alpha_\rho(s)$ and $\alpha_\sigma(s)$ Regge trajectories, 
from our constrained Regge-pole amplitudes. 
 (Right) $\alpha_\sigma(s)$ and $\alpha_\rho(s)$
in the complex plane. Beyond $s=2\,$GeV$^2$ extrapolations of our results are plotted as dotted lines. 
 Within the input pole parameter error bands, in the case of the $\sigma$, we find two types of solutions. One set (pattern-filled band)
  has a loop in the $\mbox{Im}\alpha - \mbox{Re}\alpha$ plane. The other (gray lines), having slightly higher $\alpha'$ does not form a loop. 
At low and intermediate energies, both are similar to the trajectories of the Yukawa potential $V(r)=-{\rm G} a \exp(-r/a) /r$, shown here for three different values of  G  \cite{Lovelace}. For the G=2 Yukawa curve we can estimate $a\simeq 0.5 \,$GeV$^{-1}$, following \cite{Lovelace}. This could be compared, for instance, to the S-wave $\pi\pi$ scattering length $\simeq 1.6\, $GeV$^{-1}$. }
\end{figure}

We show in the left panel of Fig.~\ref{fig:trajectories} the resulting Regge trajectories, whose parameters are given in Table \ref{aba:tbl2}. We see that the imaginary part of $\alpha_\rho(s)$ is much smaller than the real part, and that the latter grows linearly with $s$. The values for the intercept and the slope are very consistent with previous determinations such as: $\alpha_\rho(0)=0.52\pm0.02$~\cite{Pelaez:2003ky}, $\alpha_\rho(0)=0.450\pm0.005$ 
\cite{PDG}, $\alpha'_\rho\simeq 0.83\,$GeV$^{-2}$ \cite{Anisovich:2000kxa}, $\alpha'_\rho=0.9\,$GeV$^{-2}$ \cite{Pelaez:2003ky}, or $\alpha'_\rho\simeq 0.87\pm0.06$GeV$^{-2}$ \cite{Masjuan:2012gc}.

\vspace{6mm}
\begin{minipage}{\linewidth}
\centering
\captionof{table}{Parameters of the $\rho(770)$ and $f_0(500)$ Regge trajectories} \label{aba:tbl2}
\begin{tabular*}{0.7\textwidth}{@{\extracolsep{\fill} }cccc}\toprule
& $\alpha_0$ & $\alpha'$ (GeV$^{-2}$)  & \hspace{5mm}$b_0$\hspace{5mm} \\\midrule
$\rho(770)$ & $0.520\pm0.002$  & $0.902\pm0.004$ & $0.52$ \\
$f_0(500)$ &  $-0.090\,^{+\,0.004}_{-\,0.012}$ & $0.002^{+0.050}_{-0.001}$ & $0.12$ GeV$^{-2}$\\
\bottomrule
\end{tabular*}
\end{minipage}
\vspace{4mm}
  
Taking into account our approximations, and that our error bands only reflect
the uncertainty in the input pole parameters, our results 
 are in remarkable agreement with trajectories from the literature that have been obtained using very different approaches.  

 If we now examine the $f_0(500)$ trajectory we see that it is evidently nonlinear and that the slope is about two orders of magnitude smaller than that of the $\rho$ (and of other typical to quark-antiquark  resonances, {\it e.g.}\ $a_2$, $f_2$, $\pi_2$). 
 This provides strong support for a non-ordinary nature of the $\sigma$ meson. Furthermore the  
 growth of $\alpha_\sigma(s)$ is so slow that it excludes the possibility 
 that any of the known isoscalar resonances lie on its  trajectory.

Furthermore, in Fig.~\ref{fig:trajectories} we show the striking similarities
 between the $f_0(500)$ trajectory and those of Yukawa potentials
in non-relativistic scattering. Of course, our results are most reliable at low energies (thick dashed-dotted line) and the extrapolation should be interpreted cautiously. Nevertheless, our results suggest that the $f_0(500)$ looks more like a low-energy resonance of a short range potential,  {\it e.g.}\ between pions,  than a bound state of a long range confining force between a quark and an antiquark. 

In order to check that our results for the $f_0(500)$ trajectory are robust,
 we have tried to  fit the pole in~\cite{GarciaMartin:2011jx}  by fixing $\alpha'$ to a more natural value, {\it i.e.}, the one 
 for the $\rho(770)$. We have obtained a $\chi^2$ which is two orders of magnitude larger, and even worse, the resulting Regge-pole amplitude on the real axis (dotted curve in the right panel of Fig.~\ref{fig:ampl}) is at odds with the expected behavior for the $S$-wave.  
  We note that this exercise also illustrates that the large resonance width is not responsible for the fact that the $f_0(500)$
 does not follow an ordinary Regge trajectory.

\end{document}